%% file: arie.tex
\documentclass[
    ,draft            
    ,numberedheadings 
  ]
  {aipproc}

\layoutstyle{6x9}


\newcommand{\AmS}{{\protect\the\textfont2
       A\kern-.1667em\lower.5ex\hbox{M}\kern-.125emS}}
\def\be{\begin{equation}}
\def\ee{\end{equation}}
\def\bea{\begin{eqnarray}}
\def\eea{\end{eqnarray}}
\def\beas{\begin{eqnarray*}}
\def\eeas{\end{eqnarray*}}
\newcommand{\D}{\displaystyle}

\begin{document}

\title{Modeling Neutrino Quasielastic Cross Sections on Nucleons and 
Nuclei}

\author{A.~Bodek}{ address={University of Rochester, Rochester, NY} }
\author{H.~Budd}{ address={University of Rochester, Rochester, NY} }
\author{J.~Arrington}{ address={Argonne National Laboratory,
Argonne,Illinois 60439, USA} }

\begin{abstract}
We calculate the total and differential
quasielastic cross sections for neutrino and
antineutrino scattering on nucleons using
up to date fits to the nucleon elastic electromagnetic form factors 
$G_E^p$, $G_E^n$, $G_M^p$, $G_M^n$,
and $F_A$ and pseudoscalar form factors. We compare  predictions
of the  cross sections for nucelons and nuclei to experimental data.  
(Presented by Arie Bodek at CIPANP2003, New York City, NY 2003)
\end{abstract}

\maketitle


\section{Introduction}

Experimental  evidence for oscillations among the three
neutrino generations has been recently reported~\cite{Fukada_98}.
Since quasielastic (QE) scattering forms an important component of 
neutrino scattering
at low energies, we have undertaken to investigate 
QE neutrino scattering using the latest information
on nucleon form factors. 

Recent experiments at SLAC and Jefferson Lab (JLab) have 
given precise measurements of the vector electromagnetic
form factors for the proton and neutron. 
These form factors can be related to the form factors for QE
neutrino scattering by conserved vector current hypothesis, CVC. 
These more recent form factors can be used to give better predictions
for QE neutrino scattering.


The hadronic current for QE neutrino scattering is given by~\cite{Lle_72}
\begin{eqnarray*}
 \lefteqn{<p(p_2)|J_{\lambda}^+|n(p_1)>  =   }
 \nonumber \\ &
\overline{u}(p_2)\left[
  \gamma_{\lambda}F_V^1(q^2)
  +\frac{\D i\sigma_{\lambda\nu}q^{\nu}{\xi}F_V^2(q^2)}{\D 2M} 
+\gamma_{\lambda}\gamma_5F_A(q^2)
+\frac{\D q_{\lambda}\gamma_5F_P(q^2)}{\D M} \right]u(p_1),
\end{eqnarray*}
where $q=k_{\nu}-k_{\mu}$, $\xi=(\mu_p-1)-\mu_n$, and 
$M=(m_p+m_n)/2$.  Here, $\mu_p$ and $\mu_n$ are the 
proton and neutron magnetic moments.
We assume that there are no second class currents, so the scalar
form factor  $F_V^3$ and the tensor form factor $F_A^3$
need not be included. 
Using the above current, the cross section is
\begin{eqnarray*}
 { \frac{d\sigma^{\nu,~\overline{\nu}}}{dq^2} = 
  \frac{M^2G_F^2cos^2\theta_c}{8{\pi}E^2_{\nu}}\times } 
\left
[A(q^2) \mp \frac{\D (s-u)B(q^2)}{\D M^2} + \frac{\D C(q^2)(s-u)^2}{\D M^4}\right], 
\end{eqnarray*}
where
\begin{eqnarray*}
\lefteqn{A(q^2)= 
 \frac{m^2-q^2}{4M^2}\left[
    \left(4-\frac{\D q^2}{\D M^2}\right)|F_A|^2 \right.}
    \nonumber \\&
\left.  -\left(4+\frac{\D q^2}{\D M^2}\right)|F_V^1|^2
  -\frac{\D q^2}{\D M^2}|{\xi}F_V^2|^2\left(1+\frac{\D q^2}{\D 4M^2}\right) 
-\frac{\D 4q^2ReF_V^{1*}{\xi}F_V^2}{\D M^2} 
    \right],
\end{eqnarray*}
$$
B(q^2) = -\frac{q^2}{M^2}ReF_A^*(F_V^1+{\xi}F_V^2),
~~C(q^2) = \frac{1}{4}\left(|F_A|^2 + |F_V^1|^2 -
 \frac{q^2}{M^2}\left|\frac{{\xi}F_V^2}{2}\right|^2\right).
$$
\begin{figure} 
\includegraphics[width=14cm,height=8.5cm]{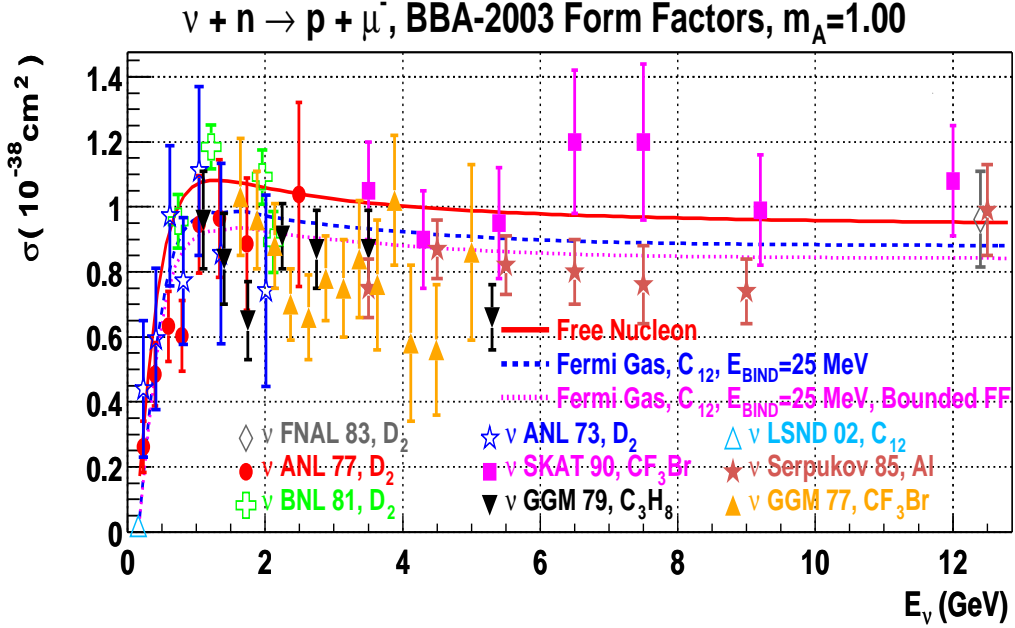}
\caption{The QE neutrino cross section along 
with data from various experiments. The calculation uses 
$M_A$=1.00 GeV, $g_A$=$-1.267$, $M_V^2$=0.71 GeV$^2$ and 
BBA-2003 Form Factors.
The solid curve uses no nuclear correction, 
while the dashed curve~\cite{Zeller_03} uses
a Fermi gas model for carbon with a 25 MeV binding energy 
and 220 Fermi momentum. The dotted curve is the prediction for Carbon
including both Fermi gas Pauli blocking and the 
 effect of 
nuclear binding on the nucleon form factors~\cite{Tsushima_03}.
The data shown~\cite{budd} are from 
FNAL 1983,
ANL 1977,
BNL 1981,
ANL 1973,
SKAT 1990,
GGM 1979,
LSND 2002,
Serpukov 1985,
and GGM 1977.}
\label{elas_JhaKJhaJ_nu}
\end{figure}

Although we have
have not shown terms of order $(m_l/M)^2$, and 
terms including $F_P(q^2)$ (which
is multiplied by  $(m_l/M)^2$), these terms
are included in our calculations~\cite{Lle_72}.)
The form factors $ F^1_V(q^2)$ and  ${\xi}F^2_V(q^2)$
are given by:
$$ F^1_V(q^2)=
\frac{G_E^V(q^2)-\frac{\D q^2}{\D 4M^2}G_M^V(q^2)}{1-\frac{\D q^2}{\D 4M^2}},
~~~{\xi}F^2_V(q^2) =\frac{G_M^V(q^2)-G_E^V(q^2)}{1-\frac{\D q^2}{\D 4M^2}}.
$$

We use the CVC to determine $ G_E^V(q^2)$ and $ G_M^V(q^2)$ 
from  the electron scattering form factors
$G_E^p(q^2)$, $G_E^n(q^2)$, $G_M^p(q^2)$, and $G_M^n(q^2)$:

$$ 
G_E^V(q^2)=G_E^p(q^2)-G_E^n(q^2), 
~~~G_M^V(q^2)=G_M^p(q^2)-G_M^n(q^2). 
$$

The axial form factor $F_A$ and the pseudoscalar form factor $F_P$
(related to $F_A$ by PCAC) are given by 
$$ F_A(q^2)=\frac{g_A}{\left(1-\frac{\D q^2}{\D M_A^2}\right)^2 },
~~F_P(q^2)=\frac{2M^2F_A(q^2)}{M_{\pi}^2-q^2}. $$
In the expression for the cross section,
$F_P(q^2)$ is multiplied by  $(m_l/M)^2$. 
Therefore, 
in muon neutrino interactions, this effect 
is very small except at very low energy, below 0.2~GeV.
 $F_A(q^2)$ needs to be extracted from 
QE neutrino scattering. At low $Q^2$,
$F_A(q^2)$ can also be extracted from pion
electroproduction data. 

Previously, people have assumed
that the vector 
form factors are 
described by the dipole approximation.
$$ G_D(q^2)=\frac{1}{\left(1-\frac{\D q^2}{\D M_V^2}\right)^2 },~~M_V^2=0.71~GeV^2$$
$$ 
G_E^p=G_D(q^2),~~~G_E^n=0,
~~~G_M^p={\mu_p}G_D(q^2),~~~ G_M^n={\mu_n}G_D(q^2).
$$
We refer to the above combination of form factors
as `Dipole Form Factors'. It is an approximation that 
has been improved by us in a previous publication~\cite{budd}. 
We use our
 updated form factors to which we refer as `BBA-2003 Form Factors' 
 (Budd, Bodek, Arrington).
We also use our updated  value~\cite{budd} of $M_A$
1.00 $\pm$ 0.020 GeV which is in good agreement with the 
theoretically corrected value 
from pion electroproduction~\cite{Bernard_01} of 1.014 $\pm$ 0.016 GeV.
\begin{figure}   
\includegraphics[width=14cm,height=8.5cm]{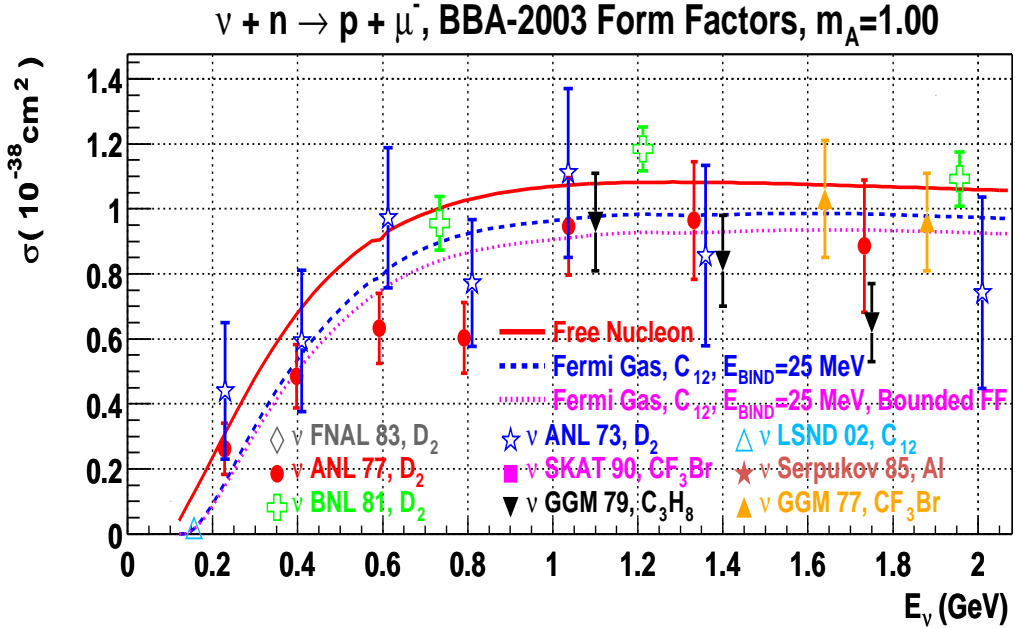}
\caption{ Same as Figure 1 
 with the $E_{\nu}$ axis limit changed to 2~GeV.}
\label{elas_JhaKJhaJ_nu2}
\end{figure}
\section{Comparison to Experimetal Data}
\begin{figure}
\includegraphics[width=14cm,height=8.5cm]{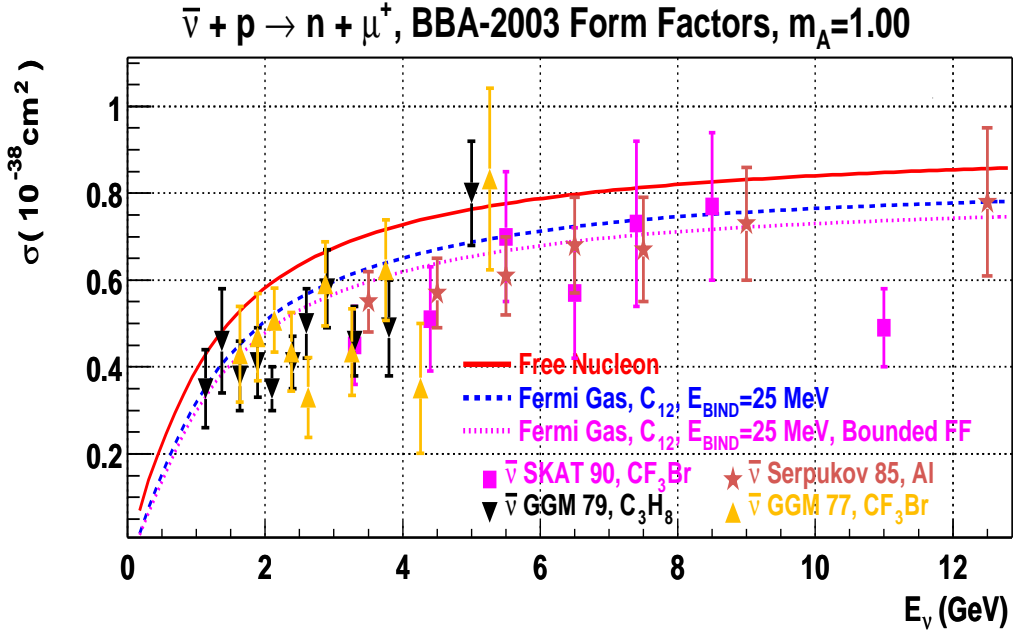}
\caption{  The QE antineutrino cross section along 
with data from various experiments.
 The calculation uses 
$M_A$=1.00 GeV, $g_A$=$-1.267$, $M_V^2$=0.71 GeV$^2$ and BBA-2003 Form Factors.
The solid curve uses no nuclear correction, 
while the dashed  curve~\cite{Zeller_03} uses
a Fermi gas model for carbon with a 25 MeV 
binding energy and 220 MeV Fermi momentum.
The dotted curve is the prediction for Carbon
including both Fermi gas Pauli blocking and the 
 effect of 
nuclear binding on the nucleon form factors~\cite{Tsushima_03}.
The data shown are from  
SKAT 1990, 
GGM 1979,
Serpukov 1985,
and GGM 1977.}
\label{elas_JhaKJhaJ_nub}
\end{figure}

\begin{figure}
\includegraphics[width=7cm,height=6.5cm]{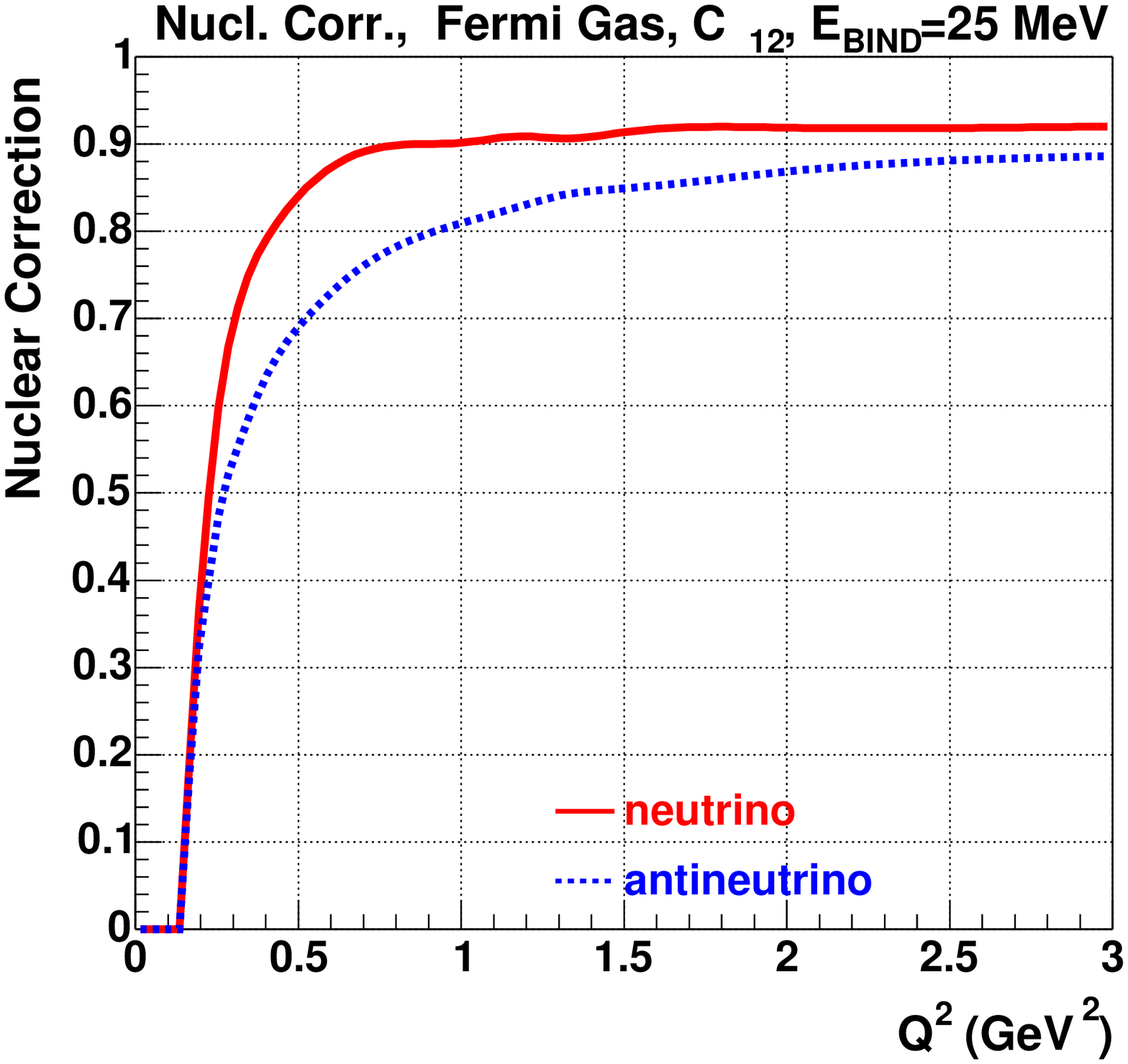}
\includegraphics[width=7cm,height=6.5cm]{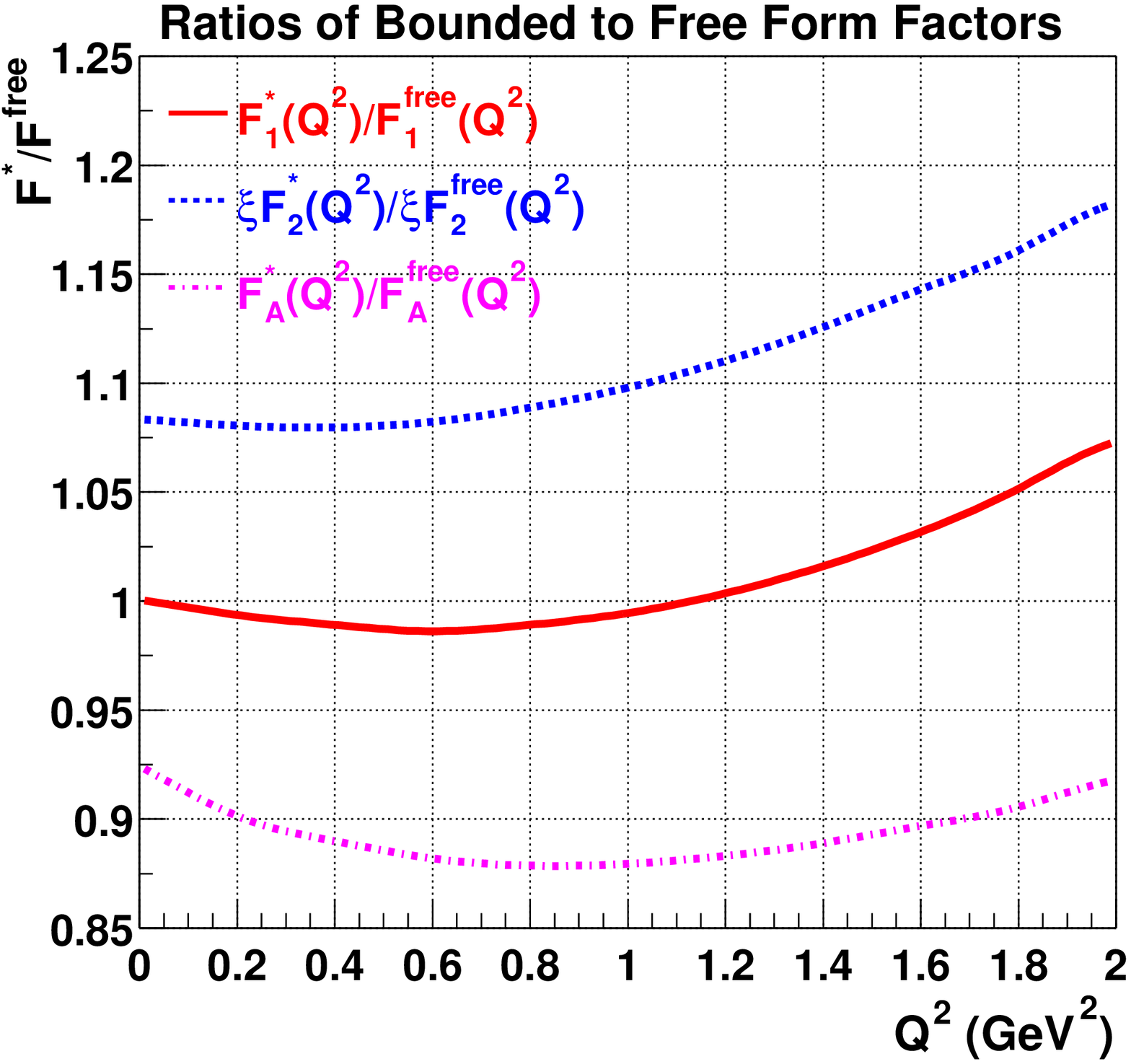}
\caption{ (a)  The Pauli blocking suppression for
a Fermi gas model for carbon with a 25 MeV 
binding energy and 220 MeV Fermi momentum. (b)
The ratio of bound to free nucleon
form factors for $F_1$, $F_2$, and $F_A$ from ref ~\cite{Tsushima_03}.
 Note that this model is only valid for $Q^2$ 
less than 1 $GeV^2$, and that the binding effects on the form factors
are expected to be very small at higher $Q^2$.}
\label{pauli}
\end{figure}

Figures~\ref{elas_JhaKJhaJ_nu},~\ref{elas_JhaKJhaJ_nu2},
and~\ref{elas_JhaKJhaJ_nub} 
show the QE cross section for 
$\nu$ and $\overline{\nu}$ with BBA-2003 Form Factors and $M_A$=1.00 GeV.
The normalization uncertainty in the data is approximately 10\%.
The solid curve uses no nuclear correction,
while the dashed curve~\cite{Zeller_03} uses
a NUANCE~\cite{Casper_02} calculation of a 
Smith and Moniz~\cite{Smith_72} based Fermi gas model
for carbon. This nuclear model includes Pauli blocking
(see Figure~\ref{pauli}(a) ) and Fermi
motion, but not final state interactions. 
The Fermi gas model was run with a 25 MeV binding energy and
220 MeV Fermi momentum.  The dotted curve is the prediction for Carbon
including both Fermi gas Pauli blocking, and the 
 effect of 
nuclear binding on the nucleon form factors
as modeled by Tsushima  {\em et al}~\cite{Tsushima_03} (see 
Figure~\ref{pauli}(b)). Note that this model is only valid for $Q^2$ 
less than 1 $GeV^2$, and that the binding effects on the form factors
are expected to be very small at higher $Q^2$.
Both the Pauli blocking and the nuclear
modifications to bound nucleon 
form factors reduce the cross section relative to the cross section with 
free nucleons. 

The updated form factors improve the agreement with neutrino QE cross section data 
and give a reasonable description of the cross sections from deuterium. 

We plan to continue to study the nuclear corrections, adopting models
which have been used in precision electron scattering measurements
from nuclei at SLAC and JLab.  For example,  we plan to study the
Pauli blocking correction using an improved Fermi
Gas model with a high momentum tail~\cite{Bodek_81}, as well as more sophisticated
nuclear spectral functions. In addition, we will continue 
to update the extraction of $M_A$ from previous neutrino experiments,
using the updated versions of the input parameters and electromagnetic form factors.

This work is supported in part by the U. S. Department of Energy, Nuclear
Physics Division, under contract W-31-109-ENG-38 (Argonne)
and High Energy Physics Division under
grant DE-FG02-91ER40685 (Rochester).

\end{document}